\documentclass[sigconf]{acmart}
\newcommand{\ignore}[1]{}

\usepackage{fancyhdr}
\usepackage{xspace}
\usepackage{graphicx}
\usepackage{subfig}
\usepackage{multirow}
\usepackage{siunitx}
\usepackage{ulem}
\usepackage{tikz}
\usepackage{mathtools}
\usepackage{enumitem}
\usepackage[compact]{titlesec}
            
\graphicspath{{figs/}}

\copyrightyear{2019} 
\acmYear{2019} 
\setcopyright{acmcopyright}
\acmConference[ISCA '19]{The 46th Annual International Symposium on Computer Architecture}{June 22--26, 2019}{Phoenix, AZ, USA}
\acmBooktitle{The 46th Annual International Symposium on Computer Architecture (ISCA '19), June 22--26, 2019, Phoenix, AZ, USA}
\acmPrice{15.00}
\acmDOI{10.1145/3307650.3322248}
\acmISBN{978-1-4503-6669-4/19/06}

\begin{document}

\title{PES: Proactive Event Scheduling for Energy-Efficient Mobile Web Computing}
\title{PES: Proactive Event Scheduling for \\Responsive and Energy-Efficient Mobile Web Computing}

\date{}
\author{Yu Feng}
\affiliation{}
\email{yfeng28@ur.rochester.edu}

\author{Yuhao Zhu}
\affiliation{}
\email{yzhu@rochester.edu}

\author{}
\affiliation{
\institution{Department of Computer Science, University of Rochester\\\texttt{http://horizon-lab.org}}}

\setlength{\textfloatsep}{6pt}
\setlength{\floatsep}{6pt}

\titlespacing*{\section}{0pt}{12pt plus 0pt minus 0pt}{6pt plus 0pt minus 0pt}
\titlespacing*{\subsection}{0pt}{8pt plus 0pt minus 0pt}{4pt plus 0pt minus 0pt}


\newcommand{\website}[1]{{\tt #1}}
\newcommand{\program}[1]{{\tt #1}}
\newcommand{\benchmark}[1]{{\it #1}}
\newcommand{\fixme}[1]{{\textcolor{red}{\textit{#1}}}}

\newcommand*\circled[2]{\tikz[baseline=(char.base)]{
            \node[shape=circle,fill=black,inner sep=1pt] (char) {\textcolor{#1}{{\footnotesize #2}}};}}

\ifx\figurename\undefined \def\figurename{Figure}\fi
\renewcommand{\figurename}{Fig.}
\renewcommand{\paragraph}[1]{\textbf{#1}~~}
\newcommand{\figline}{{\vspace*{.05in}\hline}}

\newcommand{\Sect}[1]{Sec.~\ref{#1}}
\newcommand{\Fig}[1]{Fig.~\ref{#1}}
\newcommand{\Tbl}[1]{Table~\ref{#1}}
\newcommand{\Equ}[1]{Eqn.~\ref{#1}}
\newcommand{\Apx}[1]{Appendix~\ref{#1}}

\newcommand{\specialcell}[2][c]{\begin{tabular}[#1]{@{}l@{}}#2\end{tabular}}
\newcommand{\note}[1]{\textcolor{red}{#1}}

\newcommand{\greenweb}{{\fontfamily{cmtt}\selectfont GreenWeb}\xspace}
\newcommand{\autogreen}{\textsc{AutoGreen}\xspace}
\newcommand{\pes}{\textsc{PES}\xspace}
\newcommand{\ebs}{\textsc{EBS}\xspace}

\newcommand{\RNum}[1]{\uppercase\expandafter{\romannumeral #1\relax}}


\begin{abstract}

Web applications are gradually shifting toward resource-constrained mobile devices. As a result, the Web runtime system must simultaneously address two challenges: responsiveness and energy-efficiency. Conventional Web runtime systems fall short due to their \textit{reactive} nature: they react to a user event only after it is triggered. The reactive strategy leads to local optimizations that schedule event executions one at a time, missing global optimization opportunities.

This paper proposes Proactive Event Scheduling (\pes). The key idea of \pes is to proactively anticipate future events and thereby globally coordinate scheduling decisions across events. Specifically, \pes predicts events that are likely to happen in the near future using a combination of statistical inference and application code analysis. \pes then speculatively executes future events ahead of time in a way that satisfies the QoS constraints of all the events while minimizing the global energy consumption. Fundamentally, \pes unlocks more optimization opportunities by enlarging the scheduling window, which enables coordination across both outstanding events and predicted events. Hardware measurements show that \pes reduces the QoS violation and energy consumption by 61.2\% and 26.5\%, respectively, over the Android's default \texttt{Interactive} CPU governor. It also reduces the QoS violation and energy consumption by 63.1\% and 17.9\%, respectively, compared to \ebs, a state-of-the-art reactive scheduler.

\end{abstract}

\maketitle

\section{Introduction}
\label{sec:intro}


The landscape of mobile computing has experienced a tremendous transformation over the past decade. A 2018 study shows that mobile devices have surpassed traditional devices and become the most pervasive personal computing platform~\cite{2018data}. The key enabler behind this transformation is the advancement in Web technologies, which provide a platform-independent way for mobile users to interact with
the Internet while greatly improving developers' productivity. It is estimated that over two-thirds of the US mobile traffics are contributed by Web applications~\cite{webwinning}.




Two significant but conflicting challenges stand in the way of the future mobile Web: responsiveness and energy-efficiency. Responsiveness of mobile Web applications impacts user quality-of-service (QoS), and has significant financial implications. Amazon estimates that a one-second delay in webpage load time could translate to \$1.6 billion lost in sales annually because mobile users abandon a Web service altogether if the webpage is deemed unresponsive~\cite{Eaton:2013uq}. However, a single-minded pursuit of performance to improve responsiveness is unscalable due to the tight energy budget of mobile devices, which are inherently constrained by the battery capacity without an external power supply~\cite{batterylife}.

To reconcile responsiveness with energy-efficiency, numerous prior work~\cite{Ren2017Optimise, gaudette2016improving, Zhu2013High, Zhu2015Event, zhu2016greenweb, lo2015prediction} has exploited the heterogeneous Asymmetric Chip-Multiprocessor (ACMP) architecture that has been widely adopted by today's mobile hardware
vendors such as Samsung~\cite{exynos5biglittle}, Qualcomm~\cite{kyro385}, and Apple~\cite{a11bionic}. 
The heterogeneities of ACMP, including different core
types and frequency settings, expose a large performance-energy trade-off space. Although different in design and implementation, today's ACMP schedulers share one common idea: consume ``just enough'' energy for a given responsiveness target (deadline). In particular, since mobile applications are event-driven where state transitions are triggered only by events such as user interactions, the scheduling decisions are applied at an event-granularity.

However, a fundamental limitation of existing approaches is that they are \textit{reactive} by nature in that they provision hardware resources to an event only after it has been triggered. Coupled with the reactive strategy is their \textit{localized optimizations} that schedule events one at a time without accounting for the dynamics of future events. Collectively, existing schedulers miss great optimization opportunities due to the limited event scheduling scope and the inability to coordinate scheduling decisions across events.






Our key idea is that Web runtime systems can significantly improve application responsiveness and energy-efficiency by \textit{proactively anticipating future events} and thereby \textit{globally coordinating scheduling decisions across events}. To that end, we propose Proactive Event Scheduling (\pes), which continuously predicts events that are likely to happen in the near future and coordinates event executions across both outstanding and predicted events. In particular, \pes speculatively executes future events in a way that satisfies the deadlines of all the events while minimizing the global energy consumption. Fundamentally, \pes enlarges the ``event scheduling window'' and unlocks more optimization opportunities, similar to how microarchitecture speculative techniques increase the instruction window and offer more scheduling opportunities~\cite{hennessy2018computer, kaeli2005speculative}.

Critical to the event prediction scheme in \pes is the combination of statistical inference and program analysis. Through characterizing real-world user interactions, we find that user events within an interaction session exhibit strong temporal correlation, which allows us to infer future events from past events. The accuracy of such a prediction strategy can be further improved with program analysis. The intuition is that program control flow analysis helps narrow down all possible next events mandated by the application logic, tightening the prediction space used by the statistical inference model. We propose a hybrid learning-analytical approach that accurately predicts user event sequences with low overhead.


Leveraging the predicted event sequences, we introduce a new scheduling algorithm that coordinates pending events with predicted events for energy and QoS optimizations. The scheduler wisely schedules events to different ACMP configurations to minimize the global energy consumption while satisfying individual events' QoS requirements. We find that this scheduling task can be formulated as an integer linear programming (ILP) problem, which can be efficiently computed on the fly with near-optimal solutions.

We integrate \pes with Google's open-source Chromium Web browser engine~\cite{chromium}. We evaluate \pes using the ODROID XU+E~\cite{odroidxue} development board, which contains the Exynos5410 SoC that is used in Samsung Galaxy S4 smartphone. Based on real hardware measurements, \pes achieves 26.5\% energy savings and 61.2\% QoS improvements over the Android's default \texttt{Interactive} CPU governor. \pes also achieves 17.9\% energy savings and 63.1\% QoS improvements over \ebs~\cite{Zhu2015Event}, a state-of-the-art reactive scheduler.


In summary, this paper makes the following contributions:

\begin{itemize}[topsep=4pt]
  \setlength\itemsep{0pt}
  \item We quantitatively demonstrate the inefficiencies of existing reactive schedulers for mobile applications.
  \item We propose to combine statistical inference with application analysis to predict future events. The predictor achieves high prediction accuracy while naturally adapting to different user behaviors and application contents.
  \item We introduce a new scheduling framework, \pes, that simultaneously improves responsiveness and energy-efficiency of the mobile Web applications. \pes proactively speculates user events and globally coordinates event executions based on constrained optimizations.
  \item Our evaluation results show that \pes achieves significant energy savings while reducing QoS violations compared to existing schedulers, and is closed to oracle.
\end{itemize}

The rest of the paper is organized as follows. \Sect{sec:background} describes the scope of mobile Web and introduces the background of Web runtime. \Sect{sec:exp} presents the experimental methodology. \Sect{sec:char} characterizes real mobile Web applications to motivate the need for a proactive scheduler. \Sect{sec:framework} describe the design and implementation of \pes. \Sect{sec:eval} quantifies the benefits of \pes against other existing scheduling mechanisms. \Sect{sec:related} puts \pes in the broad content of mobile (Web) optimizations. \Sect{sec:disc} discusses the limitation and future developments of \pes, and \Sect{sec:conc} concludes the paper.

\section{Background}
\label{sec:background}

We first discuss the broad scope of Web computing and introduce its fundamental event-driven execution model. We also describe how QoS is evaluated in the mobile Web.



\paragraph{Web Applications} Broadly, Web applications are applications that are developed using core Web languages including HTML, CSS, and JavaScript. Web applications not only include over 1.6 billion webpages~\cite{webpagecounts} that are accessed through Web browsers, but also ``hybrid'' applications that are internally rendered by a Web browser engine and are wrapped by a native shell~\cite{uiwebview, webview}. The combination of conventional webpages and hybrid applications accounts for a majority of today's mobile applications~\cite{html5rise}, and contributes to more than two-thirds of today's mobile Internet traffic~\cite{webwinning}.


Web applications are platform-independent in that they rely on the Web browser engine as a virtual machine or a runtime system that dynamically translates HTML, CSS, and JavaScript to re-target different mobile platforms. Although beneficial to development productivity, the dynamic translation layer introduced by the Web runtime incurs significant compute overhead. Prior work has shown that mobile Web applications demands over 80\% CPU usage~\cite{huang2012close}, which in turn dominates the energy consumption of the mobile device~\cite{mobilecpu}. Reducing the compute energy consumption while improving user-experience is thus the main goal of our work.

\paragraph{Event-Driven Execution Model} Mobile Web adopts the event-driven execution model where user interactions (e.g., tapping) are translated to application events (e.g., \texttt{touchstart}) defined by the Document
Object Model (DOM) and implemented by JavaScript. Each event is registered with an event handler (i.e., callback function) that is executed when the associated event is triggered.


\begin{figure}[t]
\centering
\includegraphics[width=\columnwidth]{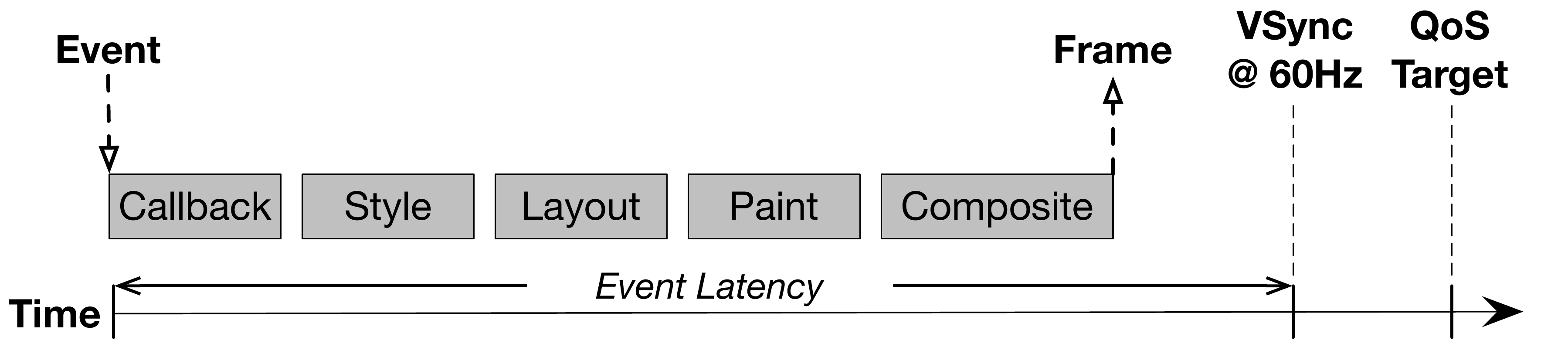}
\caption{The user QoS is dictated by the event latency, which, if exceeds the QoS target (deadline), degrades responsiveness.}
\label{fig:frame-proc}
\end{figure}

The result of an event's callback execution is then passed to the browser's rendering engine, in which each event goes through a sequence of processing stages, such as style resolution, layout, paint, and composite~\cite{renderingpipeline}, to produce a frame as a result of the user interaction. In the end, the browser submits the frame to the display upon the next display refresh, i.e., the arrival of a VSync signal~\cite{adaptivevsync}, which is mostly generated in 60~Hz on a mobile device.~\Fig{fig:frame-proc} illustrates the overall processing flow of an event. 

\paragraph{QoS Experience} A user's QoS experience is determined by the event execution latency, which is the delay between when an event (interaction) is triggered to when the corresponding frame is visualized on the display. Prior work has shown that mobile users tend to have a maximally tolerable delay of each event, also called the \textit{QoS target}~\cite{endo1996using, Zhu2015Event, yan2016redefining}. Going beyond the QoS target, would lead to an unsatisfactory QoS experience. In the example of~\Fig{fig:frame-proc}, the event latency meets the event's specified QoS target.
Note that the event latency includes an idle period between when a frame is prepared and when the display refreshes.

\section{Experimental Setup}
\label{sec:exp}


\paragraph{Software Setup} We use Google's open-source Chromium Web browser\cite{chromium} (Version 67.0.3360.0) as the experimental Web runtime. Chromium is the basis of not only the off-the-shelf Chrome browser, but also many Web runtime systems, e.g. the Android WebView~\cite{webview}.

\paragraph{Application Selection} We study a suite of 12 mobile Web applications that are previously used in similar studies~\cite{Zhu2014WebCore}. These 12 applications are ranked among the Alexa's top 25 webpages~\cite{alexa} and are representative of the top 10,000 webpages in terms of both application-inherent and hardware-dependent features based on principal component analysis.


\paragraph{Hardware Setup} We use the ODroid XU+E development board~\cite{odroidxue} as a representative mobile hardware platform. The ODroid XU+E board contains a Samsung Exynos 5410 SoC that is used in Samsung Galaxy S4 smartphone among other commercial mobile devices. The Exynos 5410 SoC contains an ACMP subsystem, which includes a high-performance, energy-hungry (big) core cluster consisting of four out-of-order Cortex A15 processors and a low-performance, energy-conserving (little) core cluster consisting of four in-order A7 processors. A15 operates between 800 MHz and 1.8 GHz at a step of 100 MHz while A7  operates between 350 MHz and 600 MHz at a step of 50 MHz. 

\paragraph{Energy Measurement} We focus on the processor energy consumption because the processor has gradually become the most significant power and energy consumer in a mobile device compared to other components such as the display and network. We leverage the build-in current sense resistors on the ODroid board to directly measure the processor power. We use the National Instruments DAQ Unit X-series 6366 to simultaneously collect voltage measurements of both the big and small CPU clusters at 1~KHz.



\section{Motivation and Characterizations}
\label{sec:char}

This section motivates the idea of proactive scheduling through systematically characterizing mobile Web applications. We first introduce the Web runtime scheduling and describe the inherent reactivity of existing scheduling schemes~(\Sect{sec:char:sched}). We then use a representative example to explain the inefficiencies of reactive scheduling mechanisms~(\Sect{sec:char:rep}), and show that the sources of inefficiencies are prevalent in mobile Web applications in general~(\Sect{sec:char:comph}).

\subsection{Web Runtime Scheduling}
\label{sec:char:sched}

The runtime scheduler is an important component of Web runtime systems. The scheduler determines how to best execute Web applications on the underlying system in order to optimize for user QoS and energy-efficiency. Conventional schedulers
leverage ``software knobs'' such as deciding the task order or throttling background activities
to give more memory resources to foreground tasks~\cite{blinksched}. Recent advancements in Web runtime schedulers are increasingly hardware-aware. In particular, this work considers hardware systems that incorporate the ACMP architecture due to their prevalence in today's mobile devices~\cite{exynos5biglittle, kyro385}. The ACMP architecture consists of multiple cores with different microarchitectures (e.g., out-of-order and in-order). Each core has a variety of frequency settings. Different core and frequency combinations expose a large latency-energy trade-off space to the runtime scheduler.

The goal of an ACMP-aware runtime scheduler is to find an ideal ACMP execution configuration (i.e., a <$core, frequency$> tuple) such that the events' QoS targets are satisfied with a minimal energy consumption. Switching off cores is not beneficial due to tiny slacks between two events thus not included here. Existing OS schedulers (e.g., the Android CPU governor~\cite{android_cpufreq}) are QoS-agnostic in that they do not take into account an event's QoS target during scheduling. Recent work has started investigating event-based, QoS-aware scheduling mechanisms that attempt to minimize energy in the presence of event QoS targets~\cite{lo2015prediction, Zhu2015Event, shingari2018dora}.

However, all existing schedulers suffer from one major inefficiency: they are \textit{reactive} by nature as they consider only events that have been triggered. As a result, they necessarily apply \textit{local} scheduling decisions in that they schedule events one at a time without considering the interferences from other events. By event interference, we refer to the fact that the execution of the current event will necessarily affect the start time of the subsequent events. Collectively, existing schedulers lack the ability to coordinate across events and miss optimization opportunities.



%

We now use one representative mobile Web application as a case-study to explain the inefficiencies of reactive schedulers. We then expand our analysis to include a comprehensive set of applications to show the general trends.

\subsection{Representative Analysis}
\label{sec:char:rep}




\begin{figure}[t]
\centering
\includegraphics[width=\columnwidth]{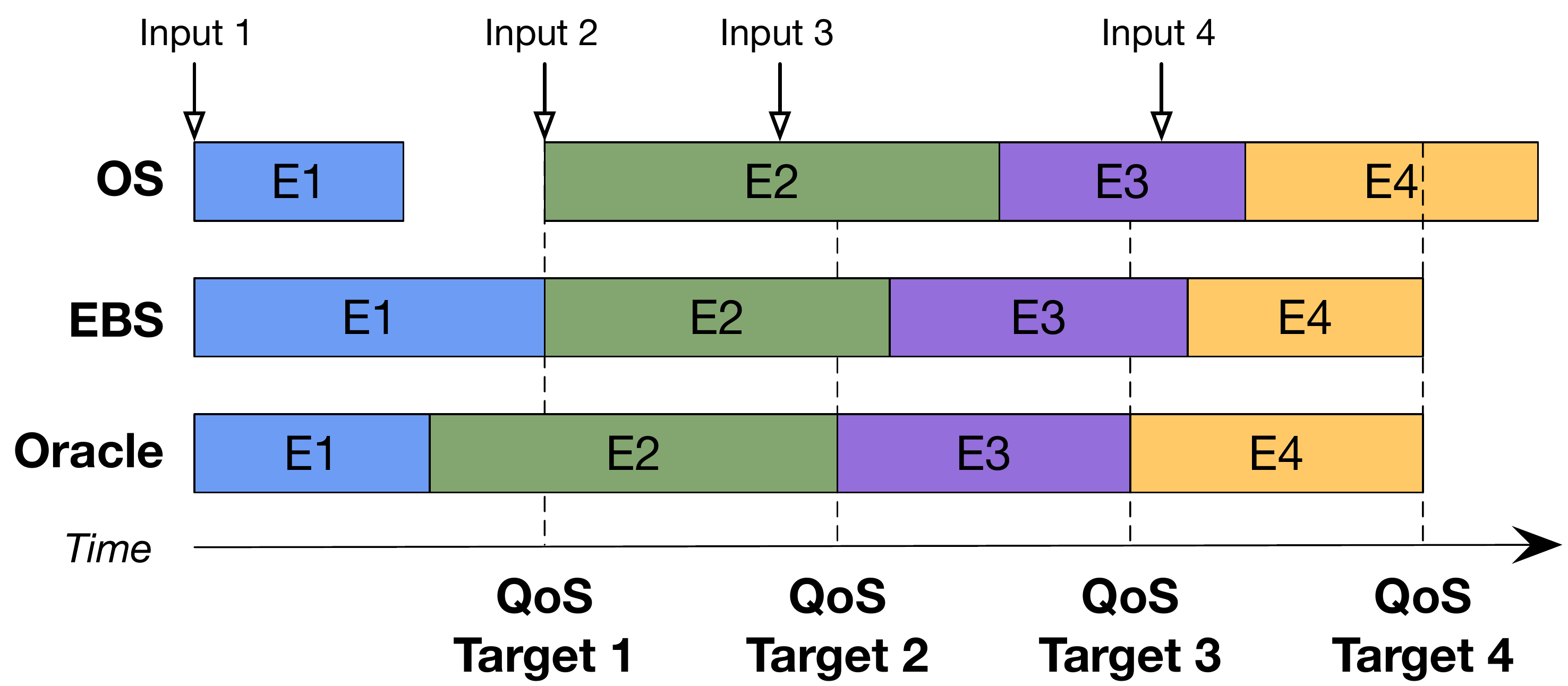}
\caption{Comparison of different scheduling mechanisms using a representative interaction sequence from~\texttt{cnn.com}. Each input (top of the figure) triggers an event execution, and corresponds to a particular QoS target (bottom of the figure). OS and \ebs are reactive schedulers and lead to QoS violations or energy waste. The oracle scheduler can proactive coordinate executions across events, and thus eliminates all QoS violations and minimizes energy consumption.} 
\label{fig:sched_example}
\end{figure}

We use a snapshot of an event sequence taken while interacting with \texttt{cnn.com} to illustrate the inefficiencies of existing schedulers. The interaction trace contains four inputs, each of which triggers an event execution. Each event has a particular QoS target (deadline) that the runtime system strives to meet in order to achieve responsiveness. We focus on the three fundamental user interactions: load, tap, and move, and use 3~s, 300~ms, and 33~ms as the QoS target for their corresponding events, respectively~\cite{zhu2016greenweb}. We abstract away the event details and use numeric representations to denote the events.

We compare three different scheduling mechanisms in~\Fig{fig:sched_example}: (1) an OS scheduler using the \texttt{Interactive} CPU governor that is QoS-agnostic, (2) \ebs which represents the state-of-the-art QoS-aware scheduler, and (3) an oracle scheduler.


\paragraph{OS Scheduler} The OS scheduler finishes the first event \textsf{E1} before the deadline. However, \textsf{E1} leaves a latency slack that could have been exploited to save energy by lowering the hardware capability. The second event \textsf{E2} misses the deadline and thus violates the user QoS requirement. This QoS violation happens because the OS scheduler does not explicitly consider QoS targets; instead, it adjusts the ACMP configurations based on the CPU utilization. \textsf{E2} has a low CPU utilization (< 70\%), and is scheduled to a low-performance configuration by the OS, which is insufficient to meet the QoS target. \textsf{E2}'s QoS violation delays the processing of \textsf{E3} and \textsf{E4}, which subsequently also miss their deadlines.


\paragraph{QoS-Aware Scheduler} Recognizing the inefficiencies in the OS scheduling, \ebs~\cite{Zhu2015Event} explicitly schedules each event under its QoS target to better optimize for responsiveness and energy-efficiency. Specifically, before executing an event \ebs predicts the optimal ACMP configuration that would meet the event's QoS target using the minimal energy.~\Fig{fig:sched_example} illustrates the improvement of \ebs over the OS scheduler. For instance, \ebs exploits the latency slack of \textsf{E1} and thus saves energy.

However, \ebs has limitations. First, \ebs misses the deadlines of \textsf{E2} and \textsf{E3}. It misses \textsf{E2}'s deadline because the inherent workload of the \textsf{E2} is so high that even the most powerful ACMP configuration could not provide enough performance. However, \textsf{E3} would have met the deadline if scheduled individually, but misses the deadline in \ebs. This is due to the interference from \textsf{E2}, which reduces \textsf{E3}'s time budget. Second, \ebs wastes energy on \textsf{E4}. This is because \textsf{E4} is delayed due to the interference of \textsf{E3}; \ebs meets the QoS target of \textsf{E4} by scheduling it to a higher-performance configuration. However, if scheduled individually \textsf{E4} could have met its deadline with lower performance and lower energy consumption.

Overall, by being QoS-aware \ebs eliminates the QoS violation of \textsf{E4} and saves energy for \textsf{E1}. However, its limited scheduling scope of outstanding events only and inability to schedule across events lead to QoS violations for \textsf{E2} and \textsf{E3} and wastes energy on \textsf{E4}.

It is worth noting that the QoS violations introduced by \ebs (i.e., \textsf{E2} and \textsf{E3}) are \textit{not} caused by the event queuing delay. We find that the average event queue length is below 2: events almost always do not wait. This is because, different from servers that could experience traffic surges, human naturally generates interactions slowly. Thus, improving the speed of the scheduler itself would have a marginal effect. In fact, \ebs has a scheduling latency $\textless$ 1~ms.

\paragraph{Oracle Scheduler}The inefficiencies of \ebs stem from its reactive nature. \ebs schedules events only after they are triggered, and therefore has a limited scheduling scope. As a comparison,~\Fig{fig:sched_example} shows the execution profile of an oracle scheduler that has \textit{a priori} knowledge of the four-event sequence. The oracle scheduler shortens the execution of \textsf{E1} to leave enough time for \textsf{E2}, which in turns allows for enough time for \textsf{E3} and \textsf{E4}. In this way, \textsf{E3} and \textsf{E4} not only meet the QoS targets, but can achieve so with lower-performance configurations that save energy. Overall, the oracle scheduler meets the QoS targets for all four events and reduces the total energy consumption by almost one-fourth compared to \ebs.

Critically, both \textsf{E2} and \textsf{E4} start executions before the corresponding inputs are triggered. Such a proactive schedule is fundamentally unattainable in reactive mechanisms such as \ebs and the OS scheduler, and indicates the potential of a proactive scheduler that anticipates future events and coordinates event executions globally.


\begin{figure}[t]
\centering
\includegraphics[width=\columnwidth]{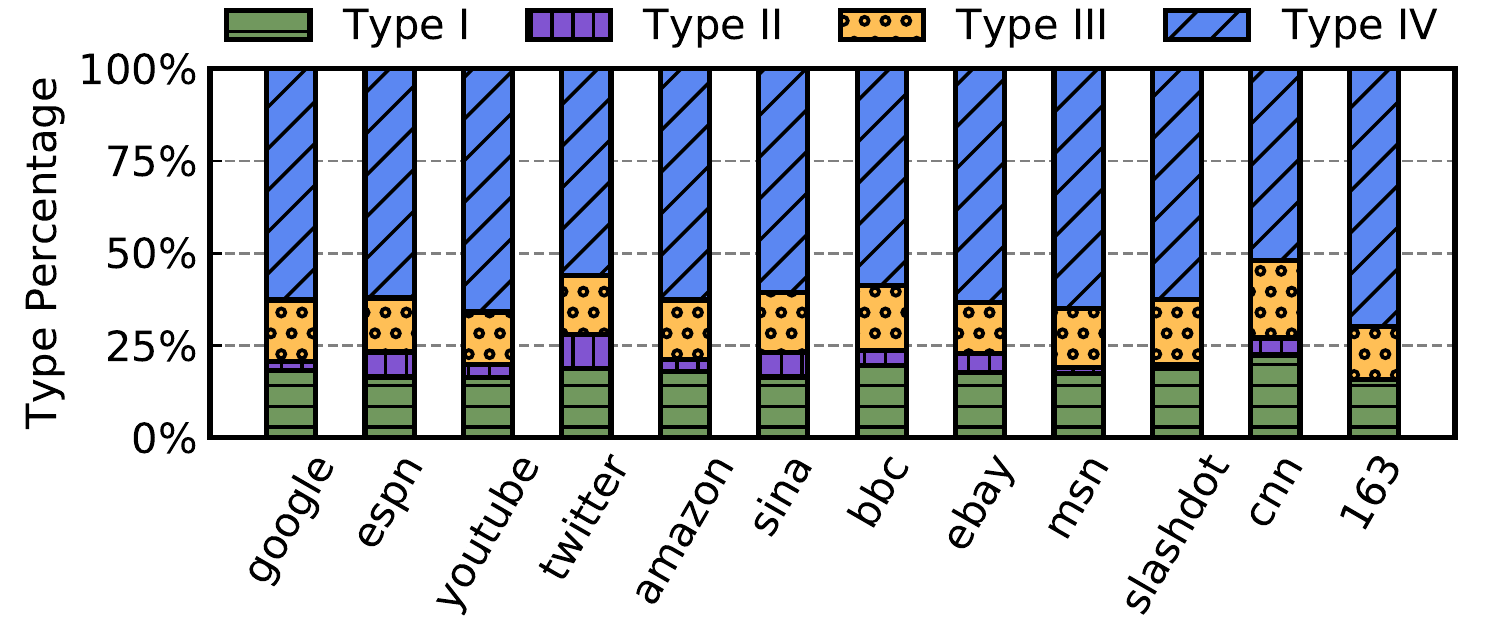}
\vspace{-15pt}
\caption{Distribution of events types under \ebs. Type I and Type II events violate QoS. Type III events meet QoS while consuming energy more than necessary. Type IV events are ``benign''; they meet QoS and provide opportunities to accommodate other types of events in a proactive scheduling.}
\label{fig:type_category}
\end{figure}


\subsection{Comprehensive Analysis}
\label{sec:char:comph}

We now expand the analysis to all 12 applications in the benchmark suite, and quantify the prevalence of the reactive schedulers' inefficiencies. To simplify discussion, we categorize events under \ebs into four categories. Note that the event categorization is not intrinsic to the events, but depends on where an event appears, which in turn reflects the scheduling policies. Our goal of event categorization is to understand the limitations of different scheduling policies, rather than the intrinsic characteristics of events.

\begin{itemize}
  \setlength\itemsep{0pt}
  \item Type I: Events whose workloads are inherently high such that even the highest-performance hardware configuration does not meet the QoS. The \textsf{E2} in~\Fig{fig:sched_example} is an example of a Type I event. For Type I events, conventional schedulers tend to consume high energy by supplying the highest-performance configuration in order to meet the deadline. However, a proactive scheduler would be able to coordinate it with its preceding events and thus meet the QoS target with lower energy.

  \item Type II: Events that could meet the deadline with a proper hardware configuration if scheduled individually, but miss the deadline at runtime due the interferences from other events. The \textsf{E3} in~\Fig{fig:sched_example} is an example of a Type II event. A proactive event scheduler would coordinate Type II events with their preceding events and thus meet the QoS targets with lower energy.

  \item Type III: Events that could meet the deadline if scheduled individually, and do meet the deadline at runtime but require higher performance than necessary due to the interferences from other events. The \textsf{E4} in~\Fig{fig:sched_example} is an example of a Type III event. A proactive event scheduler would coordinate Type III events with the interfering events so as to further exploit latency slacks to save energy.

  \item Type IV: Events that could meet the deadline with a proper hardware configuration if scheduled individually, and do not encounter interference during runtime and thus meet the QoS. The \textsf{E1} in~\Fig{fig:sched_example} is an example of a Type IV event. These events could be leveraged by a proactive scheduler to accommodate events of the previous three types for global QoS/energy optimizations.
\end{itemize}


\Fig{fig:type_category} shows how the events are distributed across the four categories. The results reveals two general trends that corroborate the observations made from the representative analysis. First, on average 21\% of the events miss the QoS target (i.e. the sum of Type I and II), and 14\% of the events potentially waste energy in meeting the QoS target (i.e. Type III). Therefore, a reactive scheduler (e.g, \ebs) does not deliver optimal results for 35\% of the events, indicating large room for improvement.


Second, the number of Type I events is almost the same as the sum of Type II and Type III events across applications. Our further investigation shows that this is because whenever a Type I event occurs, a Type II or Type III event is mostly likely to follow. The co-occurrences between Type I and Type II/III events indicate that a global scheduler that optimizes across events is likely to perform better than a local scheduler that optimizes individual events alone.

\section{Proactive Event Scheduling}
\label{sec:framework}



This section introduces \pes, a proactive event scheduler that addresses the inefficiencies in reactive schedulers. We first provide an overview of the scheduler~(\Sect{sec:framework:ov}). We then discuss the detailed \pes design, emphasizing its three components: the event predictor~(\Sect{sec:framework:pred}), the global optimizer~(\Sect{sec:framework:sched}), and the control unit~(\Sect{sec:framework:ctrl}). We provide implementation details in the end~(\Sect{sec:framework:impl}).


\begin{figure}[t]
\centering
\includegraphics[width=\columnwidth]{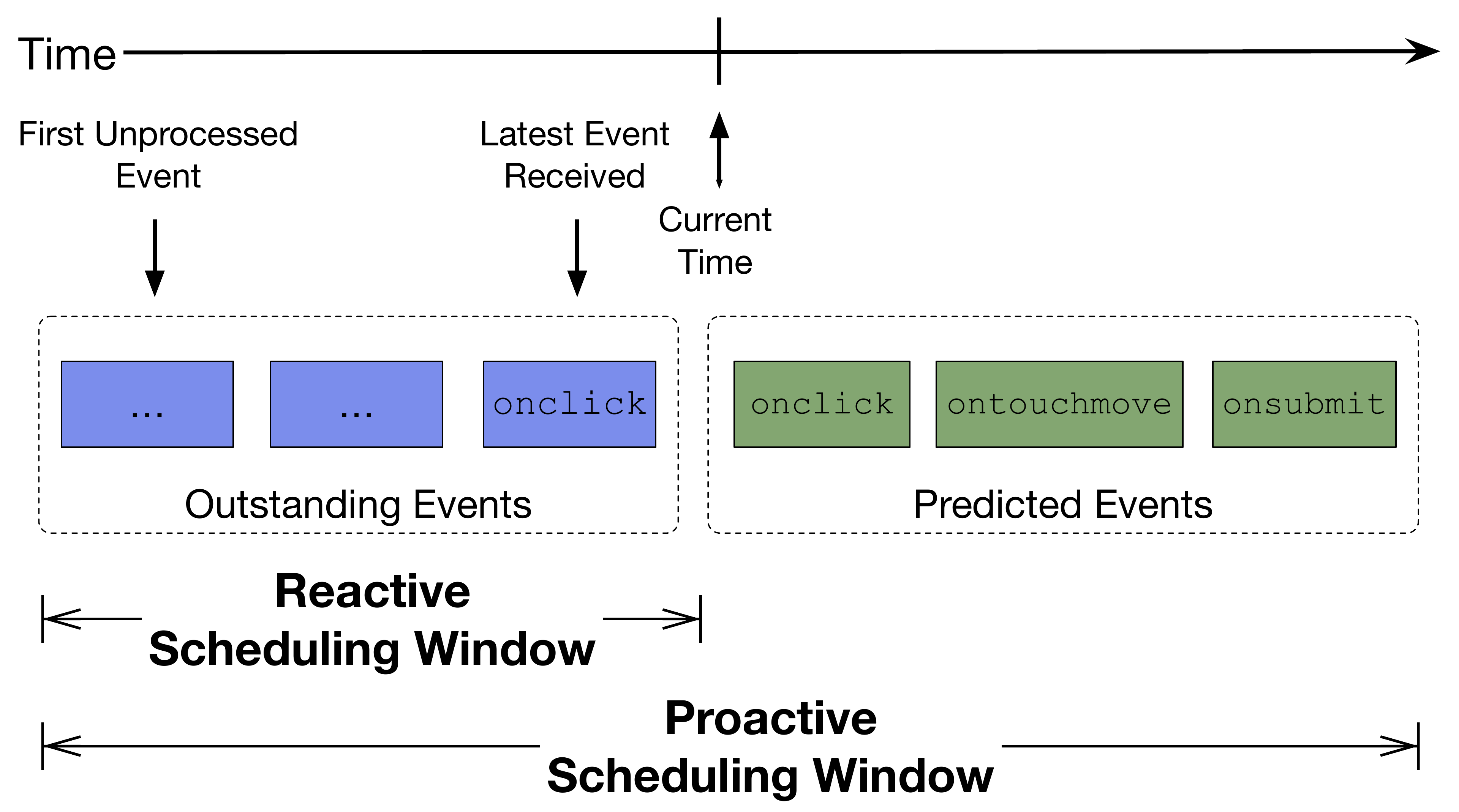}
\caption{Our proactive scheme unlocks more optimization opportunities by enlarging the scheduling window through predicting events that will likely happen in the future.}
\label{fig:exemodel}
\end{figure}

\subsection{Overview}
\label{sec:framework:ov}
We first give an intuitive illustration of the execution model under \pes, and then describe the high-level workflow of \pes.


\paragraph{Execution Model} Existing mobile Web runtime schedules events in a reactive way in that the scheduler reacts to only outstanding events, i.e., events that users have already generated but not executed yet. The key idea of \pes is to proactively anticipate future events and thereby coordinate scheduling decisions globally.

To illustrate this idea,~\Fig{fig:exemodel} compares reactive schedulers and the proposed proactive scheduler. Existing reactive schedulers has a scheduling
window limited by outstanding events that have already happened and perceived by the application. For instance in a survey section of a mobile application, a user might have triggered an \texttt{onclick} event on a checkbox, which, along with other events that have been triggered but not yet served, is scheduled by the scheduler. The scheduler optimizes for responsiveness and energy-efficiency by
leveraging the ACMP ``knobs''.

On the contrary, the proactive runtime predicts what events will likely happen in the immediate future and coordinates scheduling decisions accordingly, as illustrated in~\Fig{fig:exemodel}. For instance, it is likely that once a user clicks
a checkbox, a series of \texttt{onclick} and \texttt{ontouchmove} events will occur followed by an \texttt{onsubmit} to submit a form.
With such a prediction capability, the scheduler can take a ``sneak peek'' of future events. In this way, the proactive event scheduler not only schedule outstanding events that are waiting to be served, but also considers future events that are about to happen. \pes thus enlarges the scheduling window and unlocks more optimization opportunities.

\begin{figure}[t]
\centering
\includegraphics[width=.9\columnwidth]{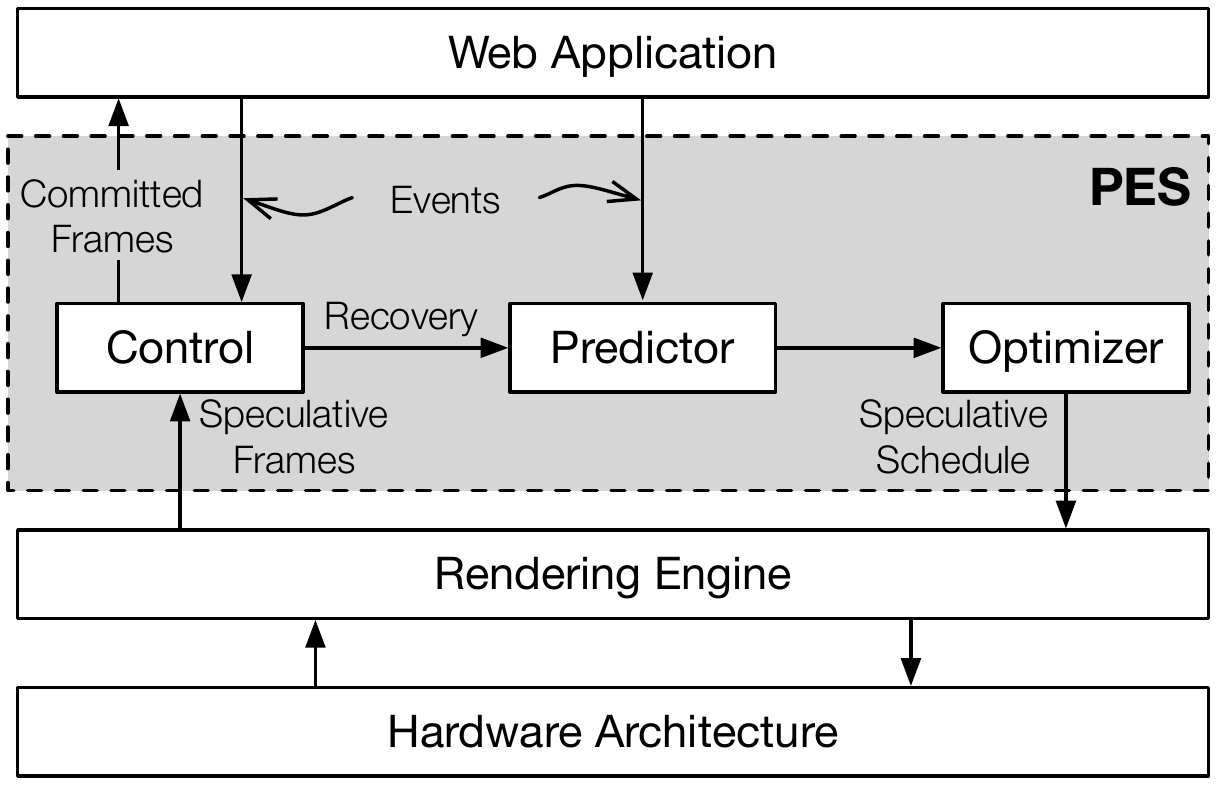}
\caption{Overview of an \pes-augmented mobile Web stack. \pes layer is shaded. See~\Fig{fig:framework} for the detailed design of \pes.}
\label{fig:overview}
\end{figure}

\begin{figure*}[t]
  \centering
  \includegraphics[width=\textwidth]{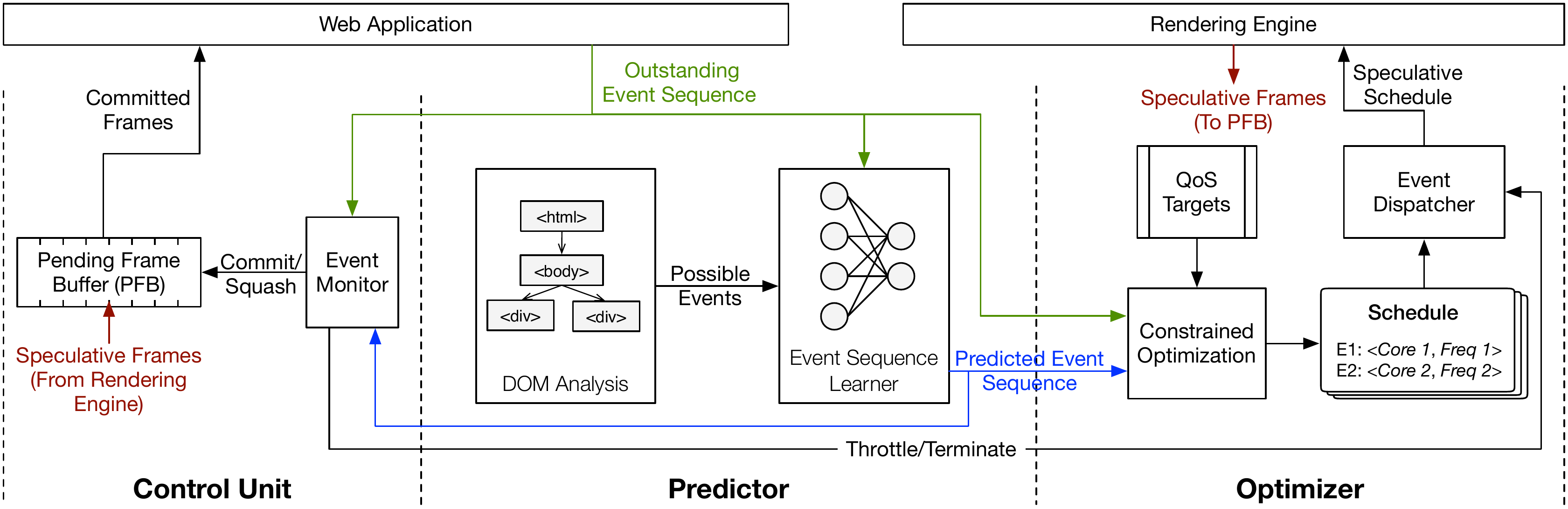}
  \vspace{-15pt}
  \caption{Detailed \pes design. The application, the \pes, the rendering engine constitute a feedback-driven optimization loop.}
  \label{fig:framework}
\end{figure*}

\paragraph{Framework Overview} In a conventional mobile Web stack, user interactions with applications (events) are forwarded to the rendering engine, which produces frames on the hardware and submits the frames to the application in reaction to the events. \pes is built on top of this architecture by adding an additional layer between the rendering engine and the application.~\Fig{fig:overview} shows an overall architecture of the \pes-augmented mobile Web stack.

The \pes layer contains three main modules, a \textit{predictor}, an \textit{optimizer}, and a \textit{control unit}. In essence, the predictor predicts a sequence of future events, which along with outstanding events are fed into the optimizer that calculates the optimal schedule, which minimizes the overall energy consumption while satisfying the QoS constraints of each event. The schedule calculated by the optimizer is in a speculative state because the predicted events have not been validated with the actual user inputs. The speculative schedule is then sent to the rendering engine, which in turn executes the schedule on the ACMP hardware to generate speculative frames.

While speculative frames are being generated, the control module monitors the actual user input events. If an actual input event matches a predict event, the controller would commit the corresponding speculated frame to the application for display; otherwise the controller drops all the remaining speculative frames, and alarms the predictor to reboot prediction. When all the speculative frames are committed, i.e., no predicted events are left, the predictor starts a new around of event sequence prediction.


\subsection{Event Predictor}
\label{sec:framework:pred}

The event predictor predicts the upcoming events during a user interactive session.~\Fig{fig:framework} shows the detailed design of \pes, where the predictor feeds the sequence of predicted events to the optimizer, and interfaces with the controller to handle mis-predictions.

The key idea of predicting future events is to combine statistical inference and program analysis. We find that user interactions constitute an event sequence that exhibits strong temporal behaviors that could be statistically inferred. Meanwhile, application-inherent logics dictate all the possible future events, tightening the prediction space.~\Fig{fig:framework} shows the interaction of these two.


\paragraph{Event Sequence Learner} Formally, the goal of the event sequence learner is to estimate the following conditional probability: $p(y_{1}, ..., y_{T'} | x_{1}, ..., x_{T})$, where $\{x_{1}, ..., x_{T}\}$ is the event sequence occurred so far, and $\{y_{1}, ..., y_{T'}\}$ is the predicted sequence. The event sequence learner operates in a recurrent fashion where every step generates a feature vector to predict the immediate next event. The predicted event is fed back to the learner to predict the subsequent event. Note that, the event sequence learner only predicts the type of immediate next event, not when it will be triggered.



We choose to construct the prediction model based on logistic regression. Specifically, the event sequence learner employs a set of logistic models, each of which estimates the probability of one possible next event through $ln(p/(1-p)) = {\textsc{\bf x} \beta}$, where $p$ is the probability that an event will be triggered, $\textsc{\bf x}$ is the feature vector, and $\beta$ are trained coefficients. In the end, the event with the highest probability is deemed the next event. There are other alternatives for temporal prediction such as Long Short-Term Memory (LSTM). However, we find that logistic regression provides sufficient accuracy with low compute overhead as we will quantify in~\Sect{sec:eval:oh}.

The predictor must adapt to different application contents and user behaviors. Therefore, we propose to construct the feature vector by considering both application-inherent information and runtime information of the current interaction sequence.~\Tbl{tab:fv} lists the specific features that we consider. In particular, we consider the runtime information within a window of the five most recent events. The combination of application-inherent and interaction-dependent features is aimed to adapt to different users and applications.
  
Each event prediction is associated with a confidence value denoted by the output of the logistic model, i.e., $p$. If the \textit{cumulative} confidence of the event sequence (i.e., the product of the confidence values of all the predicted events in the sequence) is below a threshold, the event sequence learner terminate the prediction and sends the predicted event sequence to the optimizer. The confidence threshold is a critical parameter that determines the number of consecutive events the predictor predicts ahead, which we dub \textit{prediction degree}. Intuitively, a greater prediction degree increases the scheduling window but introduces a higher chance of mis-prediction, and vice versa. We will show in~\Sect{sec:eval:robust} that \pes is largely robust against different confidence thresholds. We empirically use 70\% in our design.

\begin{table}[b]
\centering
\Huge
\caption{Model features.}
\vspace{-5pt}
\renewcommand*{\arraystretch}{1.1}
\renewcommand*{\tabcolsep}{13pt}
\resizebox{\columnwidth}{!}
{
  \begin{tabular}{ll}
  \toprule[0.15em]
  \textbf{Category}         & \textbf{Feature}\\
  \midrule[0.05em]
  \multirow{2}{*}{Application-inherent}  & Clickable region percentage in the viewport \\
  & Visible link percentage in the viewport \\
  \midrule[0.05em]
  \multirow{3}{*}{Interaction-dependent} & Distance to the previous click in the window \\
  & Number of navigations in the window \\
  & Number of scrolls in the window \\
  \bottomrule[0.15em]
  \end{tabular}
}
\label{tab:fv}
\end{table}

\paragraph{Web Application Analysis} The goal of the program analysis is to identify a set of events that could possibly be triggered, and thus narrow down the prediction space by the event sequence learner. For instance, if a button exists in an application but is not visible on the display (i.e., outside the viewport), the next user input will not trigger an \texttt{onclick} event on the button; similarly, no event will be triggered on an image if the image has no event associated with it.

We particularly focus on analyzing the DOM tree of a Web application. The DOM tree is a tree-like representation of the Web application where each node represents an application element. For instance, a submit button likely corresponds to a \texttt{button} node on the DOM tree. Each DOM node is registered with a set of events. Our DOM analyzer traverses the part of the DOM tree that is within the current viewport, and accumulates a set of events that are associated with the visible DOM nodes, which we call the \textit{Likely-Next-Event-Set} (LNES). The event sequencer learner would then predict the next event out of LNES.

The challenge of the DOM analyzer is that one event's execution might mutate the visible part of the DOM tree, and thus makes identifying the LNES for the next event difficult. Addressing this challenge is critical to enable the event sequence learner to predict multiple events consecutively so as to enlarge the scheduling space.

\begin{figure}[t]
\centering
\includegraphics[trim=0 10 0 0, width=\columnwidth]{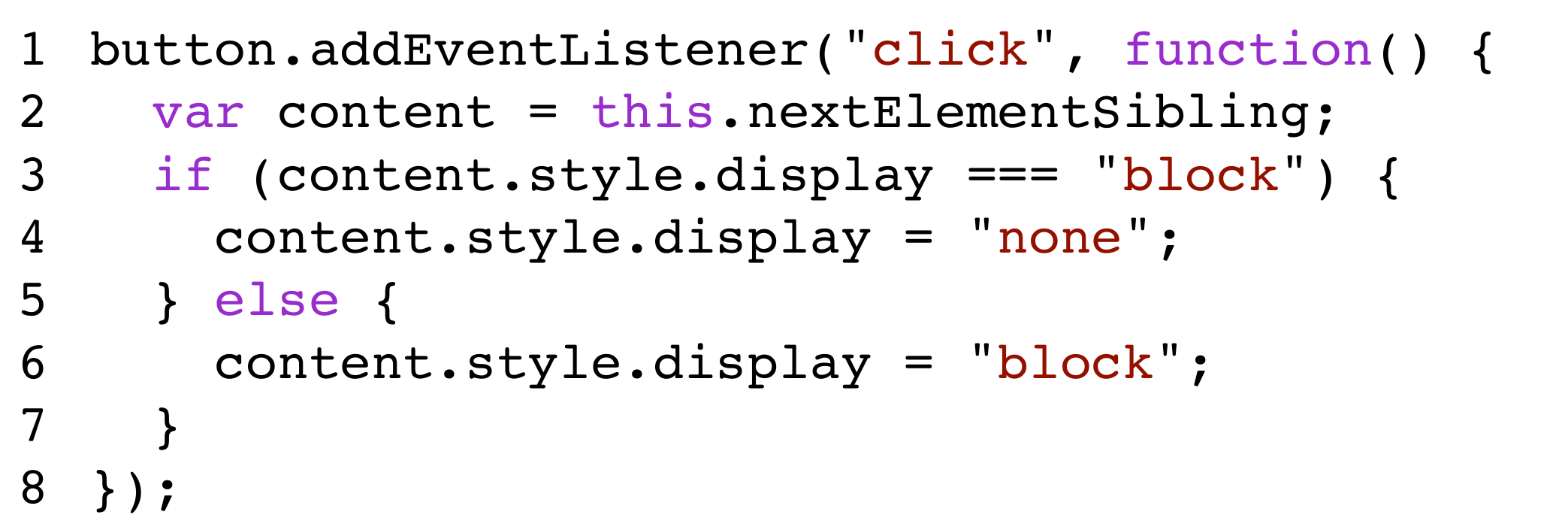}
\caption{Code snippet that toggles a collapsible menu.}
\label{fig:collapsecode}
\end{figure}

As a specific example, triggering an \texttt{onclick} event that expands a menu would show the menu and thus present more visible DOM nodes (i.e., menu items), on which more events could be triggered. \Fig{fig:collapsecode} shows a code snippet that toggles a collapsible menu where \texttt{content} is the DOM node that corresponds to the menu. The new DOM state after the~\texttt{onclick} event is not immediately clear as the callback function simply sets the display style from \texttt{none} to \texttt{block}. Fully evaluating the callback function to follow the~\texttt{content} DOM node would be a possible solution, but it defeats the purpose of globally scheduling multiple events together.

Instead, we construct a Semantic Tree during parsing, where we memoize that~\texttt{content} is a button that toggles a menu as well as the fact that the DOM node associated with the menu itself. In this way, the DOM analyzer could statically examine the Semantic Tree to identify the DOM state after the callback without having to dynamically evaluate the callback. We will show an efficient implementation of this design in~\Sect{sec:framework:impl}.

\subsection{Energy and QoS Optimizer}
\label{sec:framework:sched}

Upon receiving the predicted event sequence from the predictor, the optimizer computes a speculative schedule by combining the outstanding events with the predicted event sequence. The event dispatcher then issues the schedule to the rendering engine, which executes each event according to the schedule. We now describe schedule computation and the event dispatcher. Mis-predictions are handled by the control unit, which will be discussed in~\Sect{sec:framework:ctrl}.




\paragraph{Optimization} Intuitively, the optimization component determines the ACMP configuration (a <$core, frequency$> tuple) for each event in a way that the total energy consumption across all the scheduled events is minimized while the latency deadline (QoS target) of each event is met. We find that the scheduling task can be formulated as a constrained optimization problem, which can be efficiently solved by integer linear programming (ILP). We now describe our formulation.




We leverage the classical DVFS analytical model~\cite{xie2003compile} that estimates the execution time $T$ of a code segment:
\begin{equation}
\label{eq:1}
T = T_{mem} + N_{dep}/f
\end{equation}

\noindent where $ T_{mem} $ is the time for accessing the memory, $f$ denotes the CPU frequency, and $ N_{dep} $ is the number of CPU cycles that are not overlapped with the memory access. For the first two times an event is encountered, we measure its latency under two different frequencies and solve the system of equations as formulated by \Equ{eq:1} to obtain the values of $T_{mem}$ and $N_{dep}$. This is well-established practice used in prior work~\cite{lo2015prediction, Zhu2015Event, zhu2016greenweb}.


Each event $i$ can be scheduled to one, and only one, of the $C$ ACMP configurations. Therefore, the following ACMP configuration constraint is enforced on each event $i$:
\begin{equation}
\label{eq:2}
\sum^{C}_{j=0} \tau^{(i,j)} = 1,\ \tau^{(i,j)}\ \in\ \{0,1\}
\end{equation}

\noindent where $ \tau^{(i,j)} $ is a binary value denoting whether a particular ACMP configuration $j$ is assigned to event $i$. $ \tau^{(i,j)} $ is 1 only when the configuration $ j $ is active when executing event $ i $.

Combining \Equ{eq:1} and \Equ{eq:2}, the event latency of an event $ i $, denoted as $ \Delta t^{(i)} $, is modeled as:
\begin{equation}
\label{eq:3}
 \Delta t^{(i)} = T_{mem} + \sum^{C}_{j=0} N_{dep}/f^{(j)}\times \tau^{(i,j)},\ \tau^{(i,j)}\ \in\ \{0,1\}
\end{equation}
where $ f^{j} $ is the frequency under the configuration $j$.

In order to meet the QoS target, our formulation imposes a deadline, $ t_{c}^{(i)} $, for every event $ i $. That is:

\begin{equation}
\label{eq:4}
t^{(i-1)} + \Delta t^{(i)} \leq t_{c}^{(i)}, \forall\ i\ \in \ \{0, ..., N-1\}
\end{equation}

\noindent where $ t^{(i-1)} $ is the end time of the previous event's execution, and $ \Delta t^{(i)} $ is the latency of the current event $ i $.

The objective of the scheduler is to minimize energy consumption, which we model based on the event latency model (\Equ{eq:3}) and a power model. We construct the power model as a look-up table because the hardware exposes only a limited number of discrete frequencies. We measure the power consumption of all the frequency and core combinations offline, and persist them in a local storage file, which gets loaded into the runtime by \pes when an application boots. This is similar to the practice in prior work~\cite{Zhu2015Event}. Note that the energy consumptions in the evaluation are \textit{measured} rather than using this power estimation model.


%


Given the constraints and the power modeling, we formulate the scheduling task as an optimization problem:
\begin{align}
\label{eq:6}
min\ & \sum^{N}_{i = 0} p^{(i)} \times \Delta t^{(i)} \nonumber
\\
s.t.\ &  t^{(i-1)} + \Delta t^{(i)} \leq t_{c}^{(i)}, \forall\ i\ \in \ \{0, ..., N-1\}
\end{align}

\noindent where $ p^{(i)} $ is the power consumption of event $ i $, and $N$ is the total number of scheduled events. This optimization problem is formulated with respect to the variables $\tau^{(i,j)}$ (\Equ{eq:2}), and both the constraints and the objective are linear with respect to $\tau^{(i,j)}$.

\paragraph{Event Dispatcher} Solving the optimization problem in~\Equ{eq:6} generates a speculative schedule that assigns an ACMP configuration to each event. The event dispatcher sets up the hardware for each event based on the schedule, and then sends the event to the rendering engine. The event dispatcher would stop dispatching upon receiving a mis-prediction signal from the control unit.

One practical design decision that we take is to suppress issuing network requests before an event is confirmed to be correctly predicted. This is because network requests could have irreversible side effects (e.g., modifying server states).

\subsection{Control Unit}
\label{sec:framework:ctrl}

The control unit in \pes is responsible for validating the event prediction results from the predictor and for handling mis-predictions properly. To that end, the control unit uses a Pending Frame Buffer (PFB) to hold all the speculative frames generated from the speculative schedule. If a predicted event is confirmed to match an actual event that has occurred, the event monitor in the control unit signals the PFB to commit the corresponding speculative frame to the application for display. When all the frames are committed, the controller informs the predictor to start a new around of prediction to generate a new sequence of predicted events.

\paragraph{Handling Mis-predictions} Mis-predictions are rare as we will quantify in~\Sect{sec:eval:model}, and are very lightweight to handle when they occur. Upon an event mis-prediction, the controller simply drops all the speculative frames in the PFB, terminates the event dispatcher, and informs the event predictor to re-start the prediction. Note that the work spent on generating the frames for mis-predicted events is wasted, but the waste is minimal as we will quantify in~\Sect{sec:eval:oh}.

In addition, the control unit will disable prediction altogether if it experiences multiple ($\textgreater$ 3) mis-predictions in a row. In that case, \pes falls back to use the best reactive scheduler (i.e., \ebs in our paper). This design decision allows \pes to be robust against unexpected event behaviors.

\subsection{Implementation Details}
\label{sec:framework:impl}

\paragraph{Constrained Optimization} The optimization problem formulated in~\Equ{eq:6} can be solved by integer linear programming. We implement our own solver customized to this particular formulation instead of using a thirty-party solver such as GLOP~\cite{glop} in order to improve the runtime efficiency.

\paragraph{Constructing the Semantic Tree} We piggyback the implementation of the Semantic Tree on top of the Accessibility Tree (AT)~\cite{coreacesapi} that is widely supported in all major Web browsers such as Chrome~\cite{chromeat} and Firefox~\cite{mozillaat}.


The AT is similar to the DOM tree in structure, and reflects the semantics attributes of all the accessible nodes in the DOM tree. For instance, an AT node would tell us whether a <\texttt{div}> node is a clickable button or just a piece of text, and when click the button which other <\texttt{div}> node will become dropdown menu. In this way, by inspecting the AT the DOM analyzer could easily identify the DOM state after an event is triggered (i.e., the menu is expanded). Extending the AT for the Semantic Tree adds little runtime and implementation overhead.

\paragraph{Predictor Training} To construct the event prediction model, we record over 100 interaction traces from different users~\cite{mosaic} for the 12~applications that we study~(\Sect{sec:exp}). The traces faithfully record the timing of each event, including the user pause (thinking) time. On average, each interaction trace lasts about 110~seconds and contains about 25 total number of events (up to 70). These statistics are consistent with prior user studies in mobile computing~\cite{mobilecpu}. Our traces cover three primitive user interactions in mobile Web: loading, tapping, and moving~\cite{zhu2016greenweb}, and include different manifestations of the same interaction. For instance, our traces contain both \texttt{click} and \texttt{touchstart} events for the same ``tapping'' interaction.


The event sequence model is trained using training traces from all applications so as to be generally applicable to different applications. However, the DOM analysis at runtime naturally guides the predictor to be application-specific. We train the model offline. Training takes as little as 3 seconds on an Intel Core i5-7500 CPU at 3.40GHz, indicating the convenience of re-training if necessary. The predictive model is then integrated into Chromium Web runtime.

\section{Evaluation}
\label{sec:eval}

We first describe our evaluation methodology~(\Sect{sec:eval:method}). We then quantitatively show that the event predictor achieves high prediction accuracies, generalizes well to unseen applications~(\Sect{sec:eval:model}), and introduces negligible overhead~(\Sect{sec:eval:oh}). As a result, \pes outperforms both the Android default mechanisms as well as the state-of-the-art reactive scheduler~(\Sect{sec:eval:comp}). Finally, we conduct a sensitivity analysis to show the robustness of \pes~(\Sect{sec:eval:robust}).

\subsection{Evaluation Methodology}
\label{sec:eval:method}

\paragraph{Evaluation Benchmark} We evaluate \pes using 18 applications, which include the 12 applications used in~\Sect{sec:exp} as well as six unseen applications in order to understand the generalizability of \pes. We collect three traces for each application, and each trace is replayed under different scheduling mechanisms~\cite{mosaic}. Note that all the evaluation traces are different from the training traces used in~\Sect{sec:framework:impl} regardless of whether the applications are seen before or not. That is, we collect \textit{new} user traces for evaluation.

\paragraph{Metrics} We use two metrics to evaluate \pes: QoS violation reduction and energy savings. We define a QoS violation as an event's execution that exceeds a specified QoS target (deadline). We report the average energy consumption and QoS violation across all the events in an application.

\paragraph{Baseline} We compare against three baseline mechanisms:
\begin{itemize}
  \setlength\itemsep{4pt}
  \item \texttt{Interactive}: This is the Android's \texttt{Interactive} scheduler designed specifically to enable better interactivity. It is the default CPU governor~\cite{android_cpufreq}. It periodically samples the CPU utilization, and maximizes the CPU frequency if the CPU utilization is above 85\%.
  \item \ebs: This is the Event-based Scheduler that represents a class of reactive schedulers that optimize event executions according to their QoS targets. Before executing an event, \ebs predicts the optimal ACMP configuration that meets the event's QoS target using the minimal energy. We implement \ebs as described in Zhu et al.~\cite{Zhu2015Event}.
  \item \texttt{Oracle}: This is the oracle scheduler that has \textit{a priori} knowledge of the entire event sequence. It maximizes the energy savings during the entire application lifetime while minimizing the QoS violations.
\end{itemize}

\subsection{Event Predictor Accuracy}
\label{sec:eval:model}

\begin{figure}[t]
  \centering
  \includegraphics[width=\columnwidth]{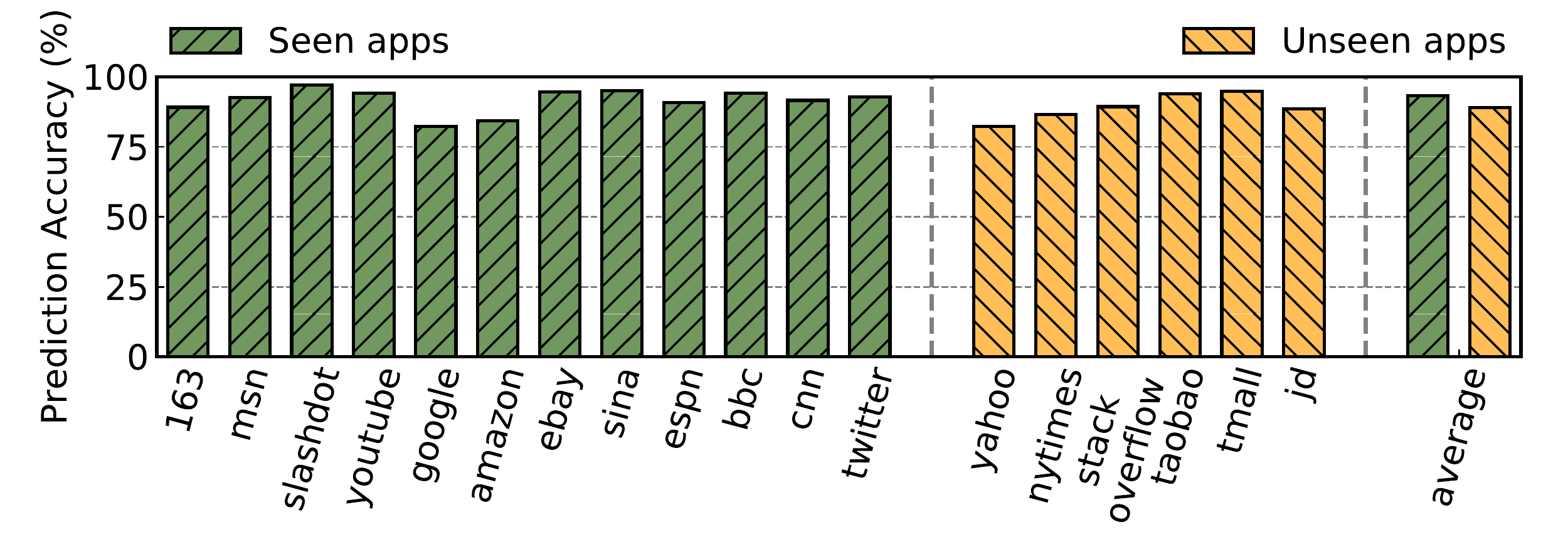}
  \vspace{-20pt}
  \caption{The event predictor accuracy. Note that all the evaluation traces are collected from new users regardless of whether the applications are seen before or not.}
  \label{fig:pred_accuracy}
\end{figure}

\Fig{fig:pred_accuracy} shows the predictor's accuracies across the 18 applications. The accuracy is the defined as the percentage of correctly predicted events. Our predictor achieves a 91.3\% prediction accuracy on average (4.1\% standard deviation) for the user interactions in the 12 seen applications. The high prediction accuracies indicate the feasibility of a simple logistic regression-based prediction model. Our predictor generalizes well to unseen applications, achieving an 89.2\% prediction accuracy (4.7\% standard deviation) for the six unseen applications. The generalizability of the predictor is a direct result of the design that augments a generic event sequence learner with application-specific DOM analyses~(\Sect{sec:framework:pred}).

\begin{figure}[t]
\vspace{0pt}
   \centering
   \captionsetup{width=\columnwidth}
   \includegraphics[trim=0 0 0 0, clip, width=.9\columnwidth]{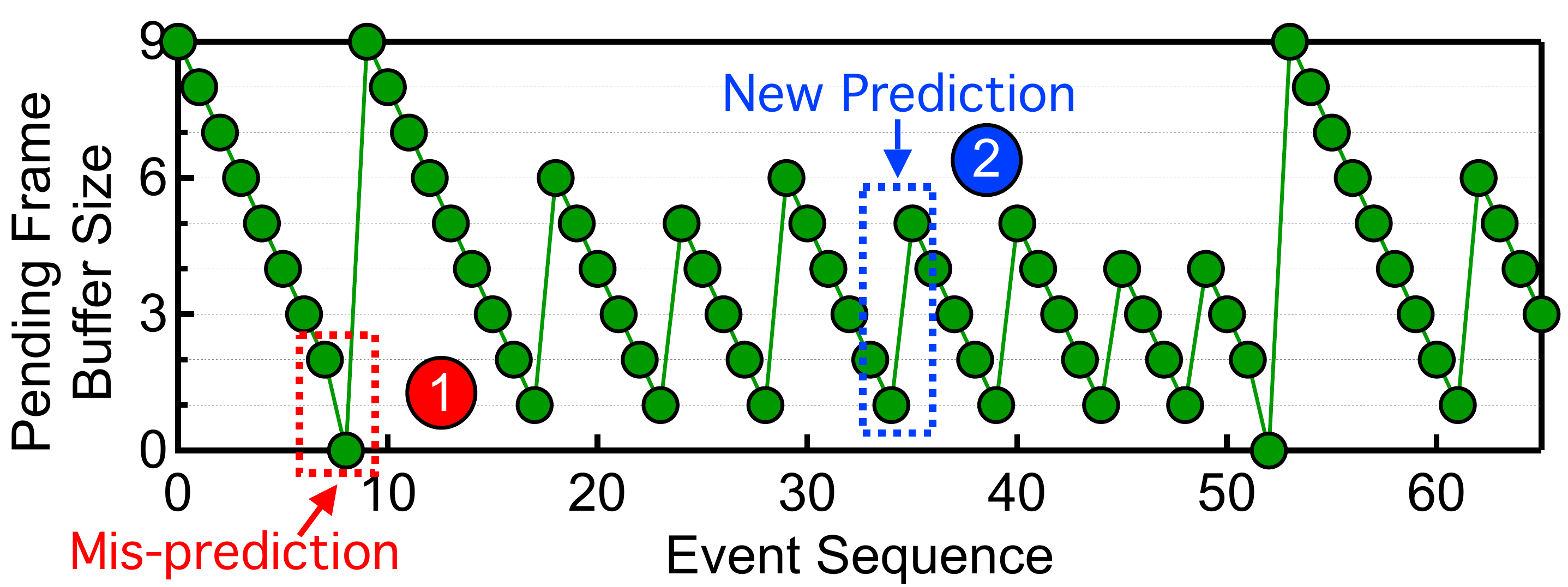}
   \vspace{-5pt}
   \caption{Pending frame buffer (PFB) size changes over time. We highlight one mis-prediction instance and one new prediction instance. Other instances are omitted due to space.}
   \label{fig:pfb}
\end{figure}

Using \texttt{ebay} as a case-study,~\Fig{fig:pfb} illustrates the dynamics of event prediction, where each <$x$, $y$> marker represents the number of speculative frames in the Pending Frame Buffer (PFB) ($y$) when a new event occurs ($x$). As described in~\Sect{sec:framework:ctrl}, when a new event occurs and is matched with a predicted event, the corresponding speculative frame will be committed and the PFB size gets decremented by 1. This is common during the application execution.

Upon a mis-prediction, all the frames in the PFB are dropped and the PFB size drops to 0, as is the case of~\circled{white}{1}. When the last predicted event is matched and the speculative frame is committed, the predictor starts a new round of prediction, and the rendering engine pushes a new set of speculative frames into PFB, as case~\circled{white}{2}.

We also observe that the prediction accuracy varies across different applications. Specifically, the accuracy varies from 97.0\% (\texttt{slashdot}) to 82.2\% (\texttt{google}). Further investigations show that the accuracy variance is mainly affected by two factors: the intrinsic properties of the application (e.g., event and DOM tree characteristics) and user interactions with the application. For instance, the prediction on applications with larger clickable area, such as \texttt{amazon}, is generally harder to predict compared to applications with less clickable applications such as \texttt{slashdot}. This indicates that an application-specific event sequence learner can potentially further improve prediction accuracy, which we leave as future work.

\subsection{Overhead Analysis}
\label{sec:eval:oh}

\paragraph{Runtime Overhead} \pes introduces three sources of addition work, all of which are negligible and are far out-weighted by the benefits of \pes. First, predicting the user events involves evaluating a simple five-variable logistic model with an overhead of about 2~$\mu$s. Second, solving the constrained optimization problem takes about 10~ms, which itself is amortized across multiple event executions. Finally, switching CPU frequencies and core migration incurs an overhead of 100~$\mu$s and 20~$\mu$s, respectively~\cite{Zhu2015Event, Zhu2013High}. The overheads are negligible compared to typical event latencies that range from hundreds of milliseconds to several seconds, and are captured by the real-system measurements.

\begin{figure}[t]
   \centering
   \captionsetup{width=\columnwidth}
   \includegraphics[trim=0 0 0 0, clip, width=\columnwidth]{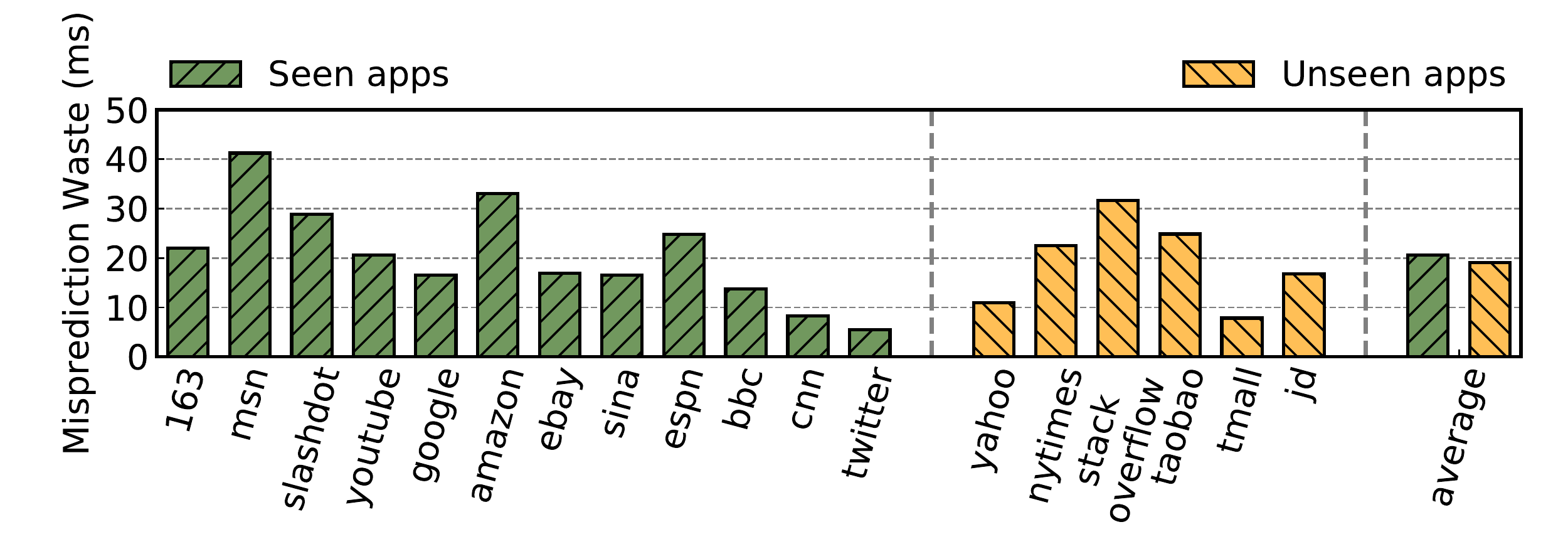}
   \vspace{-20pt}
   \caption{The average mis-prediction waste.}
   \label{fig:misprediction}
\end{figure}

\paragraph{Mis-prediction Waste} Although handling mis-prediction has almost zero cost because it involves only flushing the speculative frames, mis-predictions waste the work spent on generating the speculative frames. We define \textit{mis-prediction waste} as the time it takes to generate a speculative frame (which is eventually discarded) for a mis-predicted event. \Fig{fig:misprediction} shows the average mis-prediction waste across different applications. For both seen and unseen applications, the average mis-prediction waste is about 20~ms, which translates to an amortized waste of 2~ms per event. The average energy overhead introduced by a mis-prediction is 7.2~mJ (1.8\%) and 8.2~mJ (2.2\%) for seen and unseen applications, respectively. Combining the high prediction accuracy and the low mis-prediction waste, \pes achieves significant energy savings as we show next.



\begin{figure*}[t]
\centering
{
  \includegraphics[width=\textwidth]{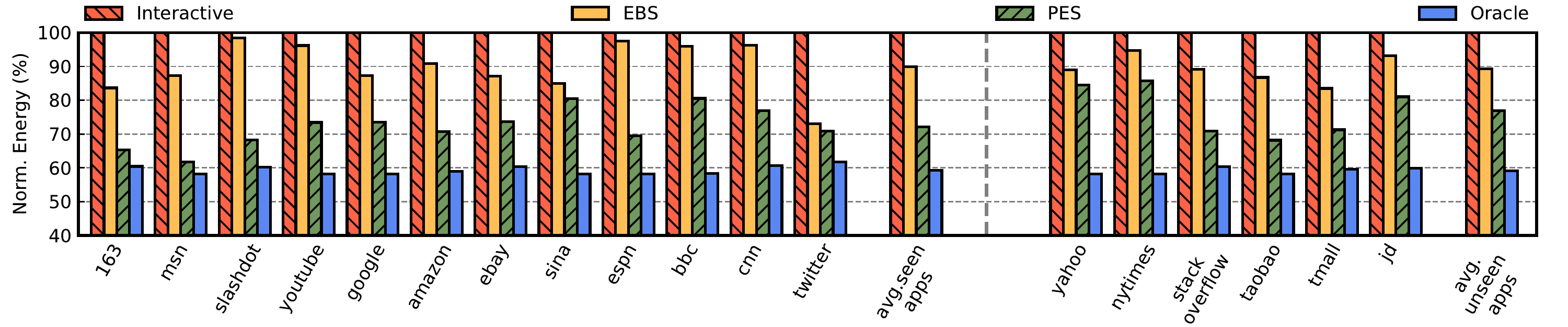}
  \vspace{-20pt}
  \caption{Energy consumption normalized to \texttt{Interactive}, which consumes the highest energy among all four schemes. Lower is better. The results of the 12 seen applications are on the right, and the six unseen applications are on the left.}
  \label{fig:compare_interactive_energy}
}
\end{figure*}

\begin{figure*}[t]
\centering
{
  \includegraphics[width=\textwidth]{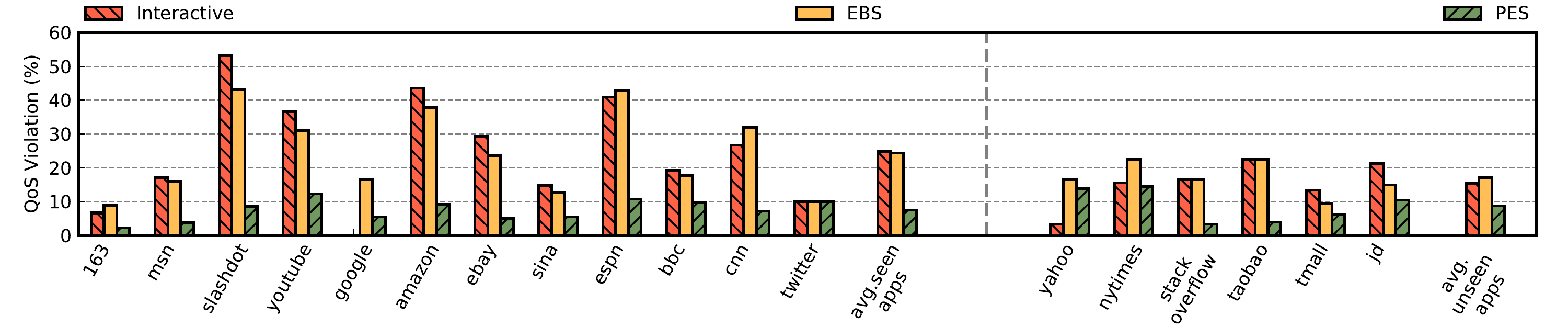}
  \vspace{-20pt}
  \caption{QoS violation. Lower is better. The \texttt{Oracle} scheduler has zero QoS violation and thus not shown. The results of 12 seen applications are shown on the right, and the results of the six unseen applications are on the left.}
  \label{fig:compare_interactive_qos}
}
\end{figure*}

\subsection{Energy and QoS Evaluations}
\label{sec:eval:comp}

\paragraph{Energy Saving} We now compare \pes with the three baseline systems described in~\Sect{sec:eval:method}.~\Fig{fig:compare_interactive_energy} shows the energy consumption of the four schemes normalized to \texttt{Interactive} for both seen applications and unseen applications. We find that \texttt{Interactive} spends over 80\% of the time running on the big core using the highest frequency, and thus consumes the highest energy. In comparison, \ebs is aware of events' QoS targets and thus is able to select a proper ACMP configuration for each event that meets the QoS target using minimal energy. On average, \ebs is able to achieve about 10.2\% energy savings compared to \texttt{Interactive}.


However, \ebs is limited by its reactive nature that schedules only events that have occurred. In contrast, \pes predicts future events to enlarge the scheduling window, and coordinates executions across events. For the 12 seen applications, \pes reduces the energy by 27.9\% and 19.8\% compared to \texttt{Interactive} and \ebs, respectively.

\pes is within 12.9\% of the energy savings of \texttt{Oracle}. The gap between \pes and \texttt{Oracle} mainly comes from two sources. First, \texttt{Oracle} can predict infinitely far ahead  (i.e., an infinite \textit{prediction degree}) because it has the knowledge of the entire event sequence, whereas \pes predicts only about five events ahead (i.e., a \textit{prediction degree} of five) before the prediction confidence drops below the threshold (See~\Sect{sec:framework:pred}). Second, \texttt{Oracle} also has a perfect prediction accuracy while \pes incurs mis-predictions as shown in~\Fig{fig:pred_accuracy}.


Interestingly, a high prediction accuracy does not always lead to a high energy saving (e.g., \texttt{sina}). This is because the prediction accuracy is not the only factor affecting energy saving. The computation intensity of the events matters too. We find that \texttt{sina} contains many compute-light events; scheduling them to a low-performance configuration leads to lower energy savings than applying the same scheduling to compute-intensive events.

To show the generazability of \pes, we also evaluate \pes under the six unseen applications. The results are shown in~\Fig{fig:compare_interactive_energy}. On average, \pes achieves  23.1\% and 13.9\% energy savings compared to \texttt{Interactive} and \ebs, respectively, both of which are slightly lower than the savings obtained from seen applications because of the slightly slower event prediction accuracy as shown in~\Fig{fig:pred_accuracy}.



\paragraph{QoS Violation} \Fig{fig:compare_interactive_qos} shows the QoS violations across the four schemes. \texttt{Oracle} completely removes the QoS violation for all the applications (and thus not shown) because it can schedule across the entire event sequence. Across the 12 seen applications, \texttt{Interactive} and \ebs incur a QoS violation at about 24.8\% and 24.4\%, respectively. In contrast, \pes decreases the QoS violation to below 10\% for most applications. On average, \pes reduces the QoS violation to only 7.5\%, indicating that \pes can simultaneously improve QoS and reduce energy compared to existing schedulers.

We also show the QoS violation reduction across the six unseen applications in \Fig{fig:compare_interactive_qos}. \pes reduce 43.7\% and 49.2\% of the QoS violations incurred in \texttt{Interactive} and \ebs. 


 
\begin{figure}[t]
   \centering
   \captionsetup{width=.9\columnwidth}
   \includegraphics[trim=0 0 0 0, clip, width=.6\columnwidth]{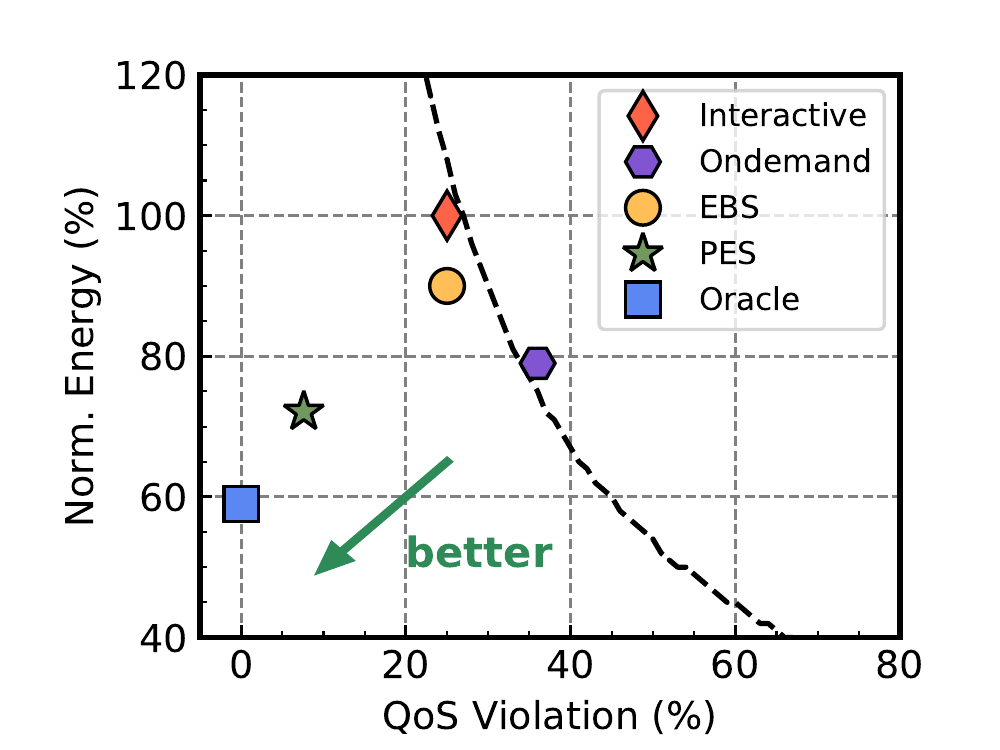}
   \vspace{-5pt}
   \caption{Pareto analysis of different scheduling mechanisms. Energy values are normalized to \texttt{Interactive}. \pes Pareto-dominates existing schemes.}
   \label{fig:pareto}
\end{figure}

\paragraph{Pareto Analysis} To summarize the benefits of \pes,~\Fig{fig:pareto} shows the Pareto-optimal frontier of all existing schemes. For the sake of completeness, we also show the results of the Android's \texttt{Ondemand} CPU governor, which favors energy savings and has much greater QoS violations, and thus is rarely used in interactive applications. \pes achieves lower energy consumption even compared to \texttt{Ondemand}. Overall, \pes Pareto-dominates all schemes.

\subsection{Sensitivity Study}
\label{sec:eval:robust}

\paragraph{Predictive Degree} We study the sensitivity of \pes with respect to one key parameter of the event predictor: the \textit{prediction degree}. Intuitively, a greater prediction degree provides more opportunities for cross-event optimizations but also introduces higher chances of mis-predictions that degrade efficiency; vice versa.

\begin{figure}[t]
\vspace{-5pt}
  \centering
  \captionsetup[subfigure]{width=0.45\columnwidth}
  \subfloat[\small{Normalized energy consumption. Lower is better.}]
  {
  \includegraphics[width=.45\columnwidth]{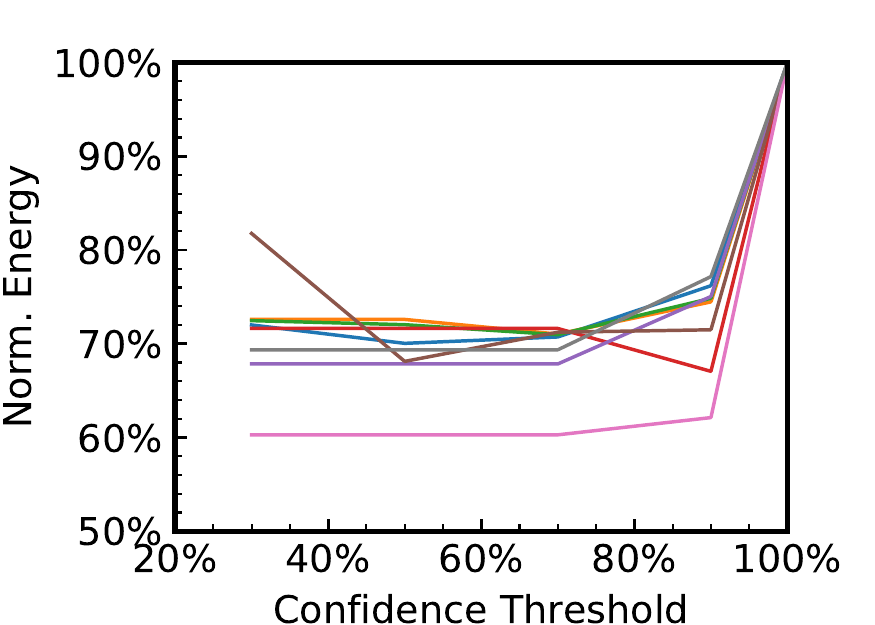}
  \label{fig:sensitivity-energy}
  }
  \hfill
  \subfloat[\small{Reduction of QoS violation. Higher is better.}]
  {
  \includegraphics[width=.45\columnwidth]{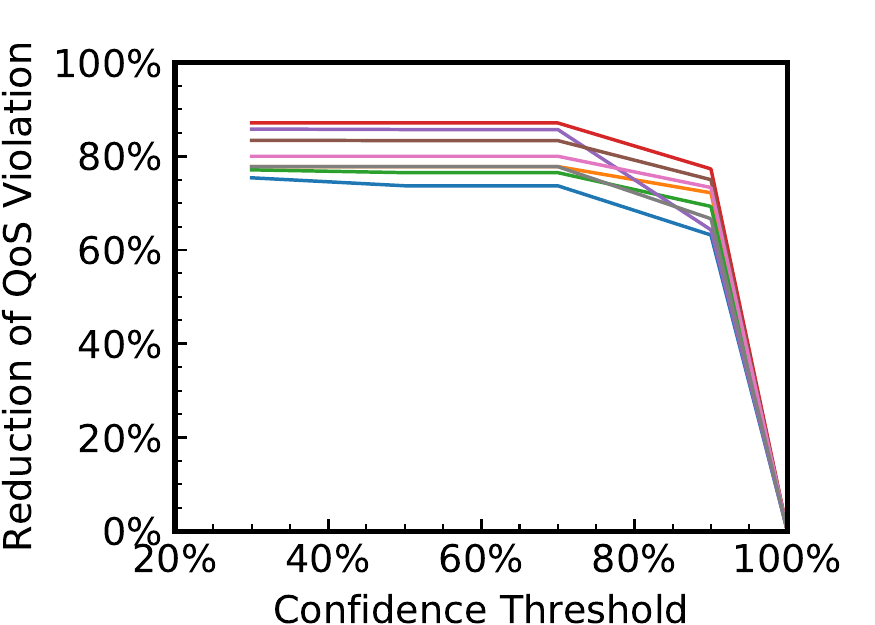}
  \label{fig:sensitivity-qos}
  }
  \vspace{-5pt}
  \caption{Sensitivity of \pes with respect to the confidence threshold. The data is normalized to \ebs. Different curves represent different applications.}
  \label{fig:sensitivity}
 \end{figure}
 
We control the prediction degree using the confidence threshold. Recall from~\Sect{sec:framework:pred} that the event sequence leaner terminates prediction if the cumulative confidence of the predicted event sequence (the product of all the predicted events' confidence values) is below a threshold. We vary the confidence threshold from 30\% to 100\%. 100\% is a confidence threshold that is too restrictive, under which predictor usually fails to make predictions and falls back to \ebs.




\Fig{fig:sensitivity} shows how the energy consumptions and QoS violation reductions change with the confidence threshold. The data is normalized to \ebs. The energy saving and QoS reduction initially improve as the threshold decreases from 100\% to 70\%. This is because a relaxed confidence threshold allows for more aggressive predictions, which enlarge the scheduling window. As the threshold further relaxes from 70\% to 30\%, the energy consumption and QoS improvements are stable because the benefits of having a large scheduling window are offset by the mis-prediction penalties. This analysis suggests that \pes is largely robust against the confidence threshold. We empirically choose 70\% as described in
~\Sect{sec:framework:pred}.

\paragraph{Predictor Design} \pes predicts future events using a combination of statistical inference (event sequence learner) and DOM analysis (\Sect{sec:framework:pred}). \pes could work with just the event sequence learner without DOM analysis, but not the other way around because DOM analysis simply identifies the current DOM state without making predictions. On average, we find that the accuracy of predicting future events without the DOM analysis  drops by about 5\%.

\paragraph{Other Devices} While we primarily evaluated \pes on the Exynos 5410 SoC, which was released in 2013, we find that \pes has similar improvements on other (more recent) mobile devices. We repeated the same experiments on the Parker SoC on Nvidia's recent TX2 board, which was released in 2017. Leveraging the DVFS capability of the Cortex A57 processors in the SoC, \pes achieves about 24.6\% energy savings compared to \texttt{Interactive}. As 75\% of today's smartphones use CPUs that are released before 2013~\cite{wu2019machine}, it is important that \pes achieves improvements on a variety of mobile devices.

\section{Related Work}
\label{sec:related}


\paragraph{ACMP Scheduling} As the underlying mobile hardware starts embracing the ACMP heterogeneous architecture, the runtime scheduler has also become ACMP-aware. Traditional ACMP schedulers such as OS governors~\cite{ondemand, android_cpufreq} are QoS-agnostic, and thus tend to miss event/job QoS targets or waste energy. Recent work has started investigating QoS-aware schedulers that make scheduling decisions based on individual event's QoS targets~\cite{Zhu2015Event, bui2015rethinking, shingari2018dora, Ren2017Optimise, Imes2015POET, lo2015prediction}. Most of such scheduling techniques are based on predicting the execution latency of the next schedulable event using history information~\cite{bui2015rethinking, Zhu2015Event}, machine learning~\cite{shingari2018dora, Zhu2013High, Ren2017Optimise}, and a combination of profiling and machine learning~\cite{lo2015prediction}. Others have studied applying control theory to ACMP scheduling~\cite{Imes2015POET, gu2008control, Imes2016Bard}.

\pes has two key distinctions. First, all existing schedulers schedule only events that have been triggered while \pes is the first work that predicts the occurrence of future events, and speculatively executes future events. Second, all existing schedulers  schedule events one at a time while \pes co-schedules outstanding events with future events to enable global QoS/energy optimizations. \pes formulates the global scheduling as an ILP problem.

\paragraph{Speculation in Interactive Applications} Prior work in mobile computing exploits speculation for various systems optimizations. Outatime~\cite{lee2015outatime} speculates user behaviors to improve the interactivity of mobile cloud games. Zare et al.~\cite{zare2016hevc} and Haynes et al.~\cite{haynes2017visualcloud} propose to predict user head movement to improve network bandwidth-efficiency in VR video streaming. Corm~\cite{mickens2010crom} speculative executes JavaScript event callbacks in order to improve the webpage loading performance. In contrast to all prior work, we propose a rigorous optimization framework that co-optimizes performance (responsiveness) and energy-efficiency at the same time.

\paragraph{Energy Optimizations in Mobile (Web) Computing} Recent work has refined the definition of QoS in mobile (Web) computing, and proposes new energy optimizations under the new QoS formulations. Gaudette et al., examines the effect of probabilistic QoS which treats QoS guarantee has a probabilistic, rather than deterministic, objective subject to uncertainties in mobile systems~\cite{gaudette2016improving, gaudette2018optimizing}. Yan et al., considers QoS variances across different users~\cite{yan2016redefining}. Our proactive scheduling mechanism is amenable to both formulations.

\pes in its current design is transparent to application developers in that it automatically optimizes for responsiveness and energy-efficiency. Prior work has demonstrated the benefits of empowering developers to better guide runtime optimizations through type systems~\cite{canino2017proactive, Sampson2011EnerJ, Cohen2012EnergyType} and program annotations~\cite{baek2010green, zhu2016greenweb}. Future work could investigate language extensions such as hints for predicting future events that could better guide \pes scheduling.

Other work has studied improving the energy-efficiency of mobile (Web) computing via hardware augmentations, which are orthogonal to our software-level runtime work. WebCore~\cite{Zhu2014WebCore} provides specialized hardware structures for key computation kernels such as CSS style resolutions and key data structures such as the DOM tree. ESP~\cite{Chadha2015Accelerating} and EFetch~\cite{chadha2014efetch} augment the hardware for efficient event processing in Web applications. Nachiappan et al. expand beyond CPU and consider
SoC-level augmentations to improve the energy efficiency of mobile (Web) applications~\cite{nachiappan2015domain, nachiappan2016vip}.

Finally, a number of work has focused on reducing the mobile Web energy optimizations through optimizing the display~\cite{Zhao2015Energy, dong2011chameleon, he2015optimizing} and radio~\cite{qian2011profiling, qian2014characterizing}. Instead, \pes focuses on optimizing the computation aspect, which has gradually dominated the energy consumption of mobile devices as radio and display technologies improve while the processor architecture becomes more power-hungry~\cite{huang2012close, mobilecpu, zhu2015role}. In our measurement of the Samsung Galaxy S4 smartphone that contains the Exynos 5410 SoC, the processor power is about 58\% of the total device power under the Interactive governor while running mobile Web applications. We thus expect that PES will lead to around 17\% total device energy reduction.


\paragraph{General Mobile Web Optimization} Prior work on mobile Web has primarily focused on improving the absolute performance through paralleling browser tasks~\cite{meyerovich2010fast, badea2010towards}, smart browser caching~\cite{zhang2010smart}, resource loading~\cite{bui2015rethinking, lymberopoulos2012pocketweb}, and improving the JavaScript engine~\cite{gal2009trace, mehrara2011dynamic}. Our work on proactive scheduling focus on user-perceivable performance (i.e., QoS), but can benefit from the absolute performance improvement techniques to better exploit time slack for global optimizations.

Another line of mobile Web research co-optimizes the mobile client with the server, either through directly augmenting the Web server or through a Web proxy~\cite{netravali2016polaris, wang2016speeding, netravali2018prophecy}. These techniques also exclusively focus on the loading phase of Web applications. \pes is a client-only solution that requires no modification to the mobile Web infrastructure, and optimizes the entire application usage session, including the loading phase as well as the post-loading interactions.



\section{Discussion}
\label{sec:disc}

\paragraph{General Applicability} While this paper focuses on one particular domain of event-driven applications, i.e., the mobile Web, the fundamental idea of proactive event scheduling is applicable to all event-driven applications. We see a broad applicability of \pes to many existing and emerging domains such as cloud services~\cite{zhu2015microarchitectural} and sensor-rich IoT systems~\cite{alam2010senaas} that are all based on the event-driven processing paradigm.

\paragraph{Other Scheduling Knobs} This paper specifically focuses on the ACMP architecture because it naturally provides a large scheduling space for performance-energy trade-offs. However, the fundamental idea of proactive scheduling is independent of the scheduling knobs, and could be applied to improve other scheduling mechanisms. For instance, dynamic display resolution scaling~\cite{he2015optimizing, dong2011chameleon} and brightness scaling~\cite{shye2009into} could be better scheduled to accommodate future events that require high user attentions.

\paragraph{Multi-application Environment} \pes is applicable to multi-programming environment where multiple applications run simultaneously. The reason is two fold. First, today's ACMP architectures provide ample hardware resources, e.g., four big and four small cores in the Exynos 5410 SoC. Therefore, \pes still has a large scheduling space even if a background application is occupying some hardware resources. Second, \pes is compatible with and can be integrated into recent proposals on interference-aware scheduling~\cite{shingari2018dora} by extending the event predictor to consider interference-sensitive features such as core utilization of co-scheduled tasks.


\section{Conclusion} 
\label{sec:conc}

To sustain the continuing growth of mobile computing, future mobile systems must compute faster and last longer. Today's mobile systems, however, are limited by their fundamental reactive nature. Reactive systems provision hardware resources to applications as demands arise, and thus are forced to apply localized optimizations. This paper promotes a proactive mobile computing platform that continuously anticipates future application events, and thereby coordinates hardware resources globally across events. We demonstrate the feasibility and benefits of a proactive system in the domain of mobile Web computing through the prototype of \pes, which simultaneously improves the energy-efficiency and the responsiveness of mobile Web applications.

\section{Acknowledgement}

We thank Brian Anderson, Ben Hayden, and Eric Seckler from the Google Chrome team and anonymous discussions from the Chromium-dev platform. This work is partially supported by a Google faculty research award. Any opinions expressed in this material do not necessarily reflect the views of the sponsors.

\raggedright
\balance
\bibliographystyle{IEEEtranS}
\bibliography{refs}
\end{document}